\documentclass[journal]{IEEEtran}
\usepackage{xcolor}
\usepackage{soul}   
\usepackage[table]{xcolor}
\usepackage{multirow}

\ifCLASSINFOpdf
   \usepackage[pdftex]{graphicx}
\else
   or other class option (dvipsone, dvipdf, if not using dvips). graphicx
\fi
\usepackage{amsmath}
\usepackage{algorithm}
\usepackage{algpseudocode}

\begin{document}

%
\title{A Distributed Consensus Algorithm for Prioritizing Autonomous Vehicle Passing at Unsignalized Intersections under Mixed Traffic}
%
%
%

\author{
Younjeong~Lee and Young~Yoon$^{*}$
\thanks{$^{*}$Corresponding author: Young Yoon (young.yoon@hongik.ac.kr).}%
\thanks{Y. Lee and Y. Yoon are with the Department of Computer Engineering, Hongik University, Seoul 04066, Korea.}%
}

\maketitle
\begin{abstract}

We propose a methodology for connected autonomous vehicles (CAVs) to determine their passing priority at unsignalized intersections where they coexist with human-driven vehicles (HVs). Assuming that CAVs can perceive the entry order of surrounding vehicles using computer vision technology and are capable of avoiding collisions, we introduce a voting-based distributed consensus algorithm inspired by Raft to resolve tie-breaking among simultaneously arriving CAVs. The algorithm is structured around the candidate and leader election processes and incorporates a minimal consensus quorum to ensure both safety and liveness among CAVs under typical asynchronous communication conditions. 
Assuming CAVs to be SAE (Society of Automotive Engineers) Level-4 or higher autonomous vehicles, we implemented the proposed distributed consensus algorithm using gRPC. By adjusting variables such as the CAV-to-HV ratio, intersection scale, and the processing time of computer vision modules, we demonstrated that stable consensus can be achieved even under mixed-traffic conditions involving HVs without adequate functionalities to interact with CAVs. Experimental results show that the proposed algorithm reached consensus at a typical unsignalized four-way, two-lane intersection in approximately 30-40 ms on average. A secondary vision-based system is employed to complete the crossing priorities based on the recognized lexicographical order of the license plate numbers in case the consensus procedure times out on an unreliable vehicle-to-vehicle communication network. The significance of this study lies in its ability to improve traffic flow at unsignalized intersections by enabling rapid determination of passing priority through distributed consensus even under mixed traffic with faulty vehicles. 

\end{abstract}


%
\IEEEpeerreviewmaketitle

\section{Introduction}\label{introduction}
\IEEEPARstart{R}{esearch} and development on autonomous vehicles is currently progressing rapidly, with the goal of commercializing fully autonomous driving technology at Level 4 or higher as defined by the Society of Automotive Engineers (SAE) in the United States~\cite{lee2024typical, barabas2017current}. Furthermore, technologies for Connected Autonomous Vehicles (CAVs) are becoming essential beyond stand-alone autonomous driving~\cite{rana2023connected}.

CAV refers to a vehicle that enables safer and more efficient driving through real-time data exchange based on V2X (Vehicle-to-Everything) communication as well as autonomous driving functions~\cite{gehrig1999dead}. Researchers are paying attention to safety and efficiency in complex traffic environments where CAVs and human-driven
vehicles (HVs) coexist~\cite{yu2025hierarchical, hu2024critical, li2023survey, ding2025ramp}. They are actively conducting research on autonomous driving operation methods under various challenging traffic conditions such as unsignalized intersections, bad weather, blind spots, pedestrian crosswalks, and night driving~\cite{SustainableMobilitySeoul2021}. Among them, unsignalized intersections are emerging as the most dangerous and complex points in the road system, and as the number of vehicles passing through the intersection increases, congestion and the risk of accidents are continuously increasing~\cite{qin2018improving,namazi2019intelligent}. In an environment without traffic lights, delays and inefficiencies inevitably occur at intersections because of the heavy reliance on drivers’ complex judgments and decisions must be relied on entirely. In particular, in situations where CAVs and HVs are mixed, the traffic congestion is expected to be aggravated due to the HV’s reaction speed and decision-making delay~\cite{Shao2023PassingMA}.

\begin{figure}[h]
    \centering
    \includegraphics[width=0.45\textwidth]{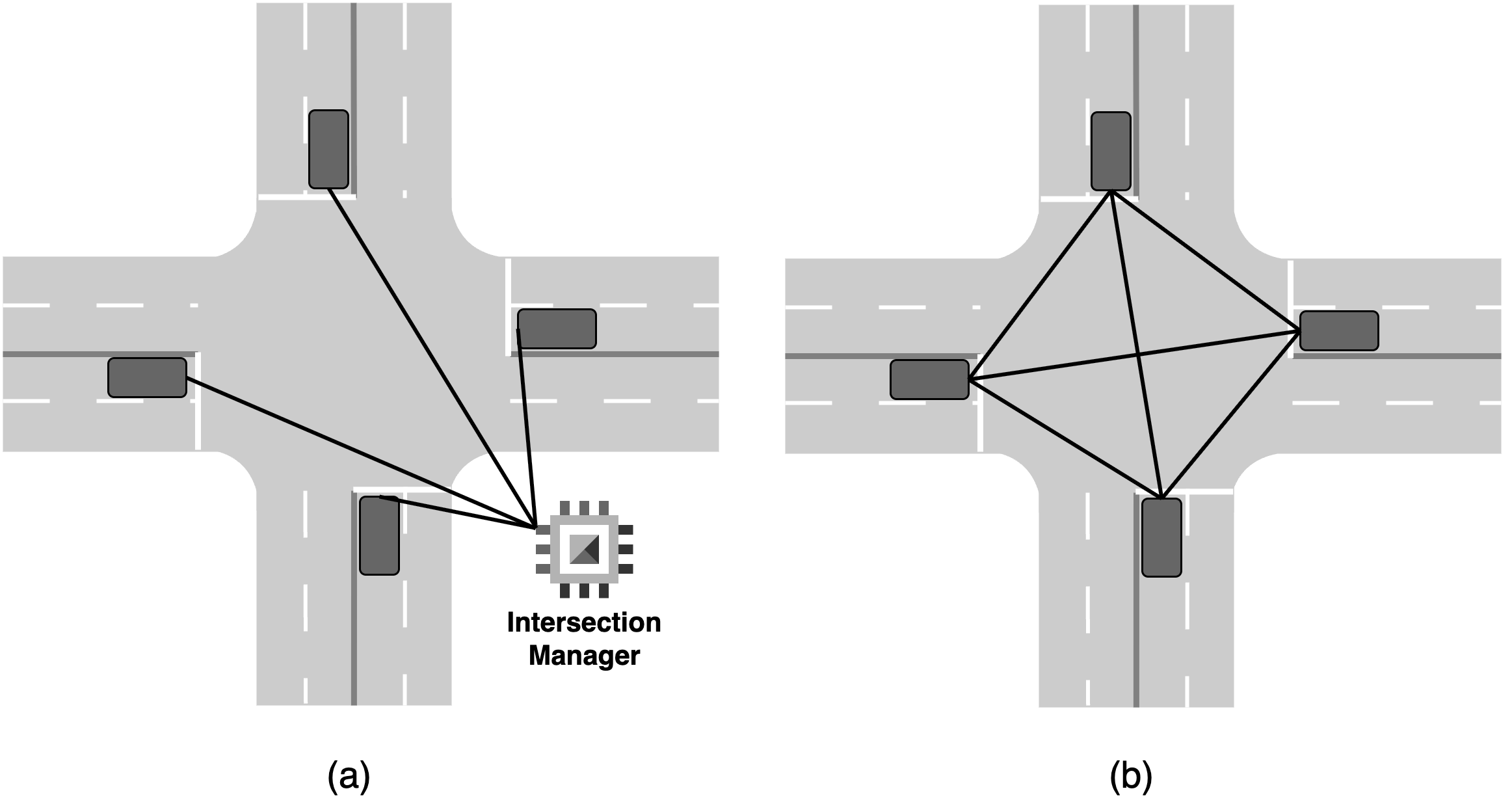}
    \caption{The communication structure of AIM: (a) Centralized AIM, (b) Distributed AIM.}
    \label{fig:communication structure of AIM}
\end{figure}

To address efficient intersection management and congestion issues at unsignalized intersections involving CAVs, Autonomous Intersection Management (AIM) techniques have emerged. AIM is a method among cooperating vehicles to improve traffic flow by minimizing stop and delay times compared to traditional traffic signal systems~\cite{zhong2020autonomous}. As illustrated in Fig.\ref{fig:communication structure of AIM}, AIMs are classified into centralized and distributed types depending on the traffic coordination approach~\cite{chen2015cooperative}.

Centralized AIM uses information collected from road infrastructure and inter-vehicle communication to allow a coordination unit, such as an intersection manager or a traffic light, to decide vehicle entry orders and wait time~\cite{dresner2008multiagent, guo2003study, wu2012cooperative}. On the other hand, distributed AIM allows each vehicle to make decisions individually through a vehicular ad hoc network, e.g., VANET (Vehicular Ad Hoc Network), without relying on a central control unit~\cite{liu2017distributed, buzachis2019distributed, xu2018distributed, cederle2024distributed}. In this case, each vehicle coordinates the entry order at intersections through interactions with surrounding vehicles, allowing for relatively fast and flexible decision-making. In distributed AIM, intersection traffic may also be temporarily coordinated by a leader selected through negotiation among vehicles in the absence of a fixed control unit.



However, neither of the two mechanisms completely resolves the conflicts and problems at intersections. For example, it is difficult to resolve deadlock situations caused by \emph{simultaneous} entry at an unsignalized intersection~\cite{liu2017distributed}. Moreover, CAVs are prone to communication failures~\cite{elliott2019recent}. 
Previous mixed-traffic intersection studies have pointed out the uncontrollability of HVs and the response-time issues of CAVs, and have proposed new control mechanisms accordingly. However, most of these studies still address the problem under the assumption that CAVs can be fully controlled~\cite{li2023survey}. As mixed-traffic conditions with partial CAV penetration are expected to remain the norm for the next several decades~\cite{namazi2019intelligent}, it is necessary to develop intersection control mechanisms that operate reliably under these mixed-traffic conditions, reflecting the practical limitations of real-world CAVs.

In particular, situations where multiple vehicles enter an intersection simultaneously increase the risk of collisions and can lead to congestion. Although effectively addressing this issue is crucial, most existing studies arbitrarily pre-assigned priority based on the direction of vehicle entry (e.g., giving priority to right-turning vehicles~\cite{vanmiddlesworth2008replacing} or to east-west vehicles~\cite{gholamhosseinian2024cai}), or designed reservation systems to prevent collisions and maintain efficient routes and speeds~\cite{hausknecht2011autonomous, abbas2023autonomous}. However, these studies did not adequately consider potential CAV failures, such as network delay and omission. These studies are not pragmatic, as they overlook the issues with the presence of HVs.

Therefore, unlike previous studies, we focus on resolving deadlock situations that arise from simultaneous vehicle entries at unsignalized intersections. Instead of adopting a centralized approach, which incurs significant infrastructure costs~\cite{au2015autonomous}, we employ a distributed approach that enables real-time vehicle-to-vehicle (V2V) communication. Considering mixed-traffic conditions involving HVs and potential non-responsive CAVs, we implement a voting-based consensus algorithm rather than assigning arbitrary priority.

The voting-based consensus algorithm is inspired by distributed consensus techniques used in networked computing systems. Unlike blockchain-based consensus algorithms involving time-consuming Proof of Work (PoW) or Proof of Stake (PoS) procedures~\cite{liao2021digital}, our approach is tailored to unsignalized intersection environments by designing a real-time voting system. The proposed system maintains partial autonomy even in the event of failures and comprehensively considers two critical properties in distributed systems—\emph{safety} and \emph{liveness}~\cite{castro2002practical}. Safety ensures that all non-faulty nodes make the same decision during the consensus process, guaranteeing consistent outcomes. Liveness ensures that the consensus process eventually terminates and reaches a final decision, thus enabling the system to proceed without indefinite delays. In real-time environments such as traffic intersections, our design aims to ensure decision consistency and enable prompt consensus, thereby supporting safe and efficient traffic flow for autonomous vehicles.


By enabling rapid right-of-way determination through inter-vehicle voting, our method addresses potential traffic stagnation in autonomous vehicle operations. While assuming an environment equipped with SAE Level-4 or higher CAVs and smart city infrastructure supporting V2V communication, the proposed system is designed to operate even without centralized control infrastructure and reflects realistic scenarios involving mixed traffic with both CAVs and HVs. The main contributions of this study are as follows.

\vspace{0.3em} 
\begin{enumerate} 
    \item Proposal of an efficient voting-based consensus mechanism for vehicles simultaneously entering an unsignalized intersection
    \item Presentation of a practical consensus approach in an environment where autonomous and human-driven vehicles coexist
    \item Implementation of a V2V communication-based consensus algorithm that achieves high throughput and rapid consensus in a distributed environment
\end{enumerate} 
\vspace{0.3em}

The remainder of this paper is structured as follows: Section II introduces related studies; Section III presents the structure of the proposed model based on its design and algorithm; In Section IV, we explain the implementation details and discuss experimental results of the algorithm across various scenarios; Section V summarizes the research and provides conclusions along with future research directions.
        \section{Related Works}\label{related_work}
        
        \subsection{Intersection Entry Priority}
        Existing studies have addressed the negotiation of vehicle entry priority in AIM using the First-Come-First-Served (FCFS) policy, which has been widely studied because it is simple to implement within the AIM framework~\cite{fajardo2011automated, li2013modeling}. In addition to FCFS, crossing priorities have been determined through intersection auctions or time value comparisons~\cite{denant2019economic, carlino2013auction, levin2016optimizing}. 
        Although various approaches such as FCFS, priority trees, and direction- or regulation-based rules have been explored, they generally overlooked simultaneous entry or equal priority situations~\cite{wang2021scheduling, dongxin2021priority, vanmiddlesworth2008replacing, gholamhosseinian2024cai, aksjonov2021rule, lu2014rule}.
        

        To the best of our knowledge, while a number of studies have addressed mixed traffic environments with both CAVs and HVs, no prior work has specifically considered the simultaneous-entry deadlock problem at unsignalized intersections under mixed traffic conditions.
        Additionally, while CAVs offer the potential to enhance traffic flow and increase road capacity through consistent and efficient operations, challenges remain due to the inherent limitations of unreliable wireless communication, such as non-deterministic message transmission delays, interference from surrounding infrastructure, and packet loss. These issues are particularly prominent in V2V communication, where heterogeneous time delays are inevitable due to differences in vehicle positions and communication hardware~\cite{li2021car, gupta2021taxonomy}. Existing studies fell short in exploring these failure scenarios. 
        
        Therefore, beyond simple priority adjustments, it is essential to consider both the communication issues of CAVs and the coexistence of HVs. It is also imperative to develop a method that can effectively manage the entry sequence of simultaneously entering vehicles with equal priority.
        
        \subsection{Distributed Consensus in CAV Environments}
        Centralized approaches face increasing management costs and failure issues as connected devices grow, pushing research toward low-latency distributed methods that take advantage of individual CAV capabilities~\cite{liu2019mathsf,sandner2020convergence,fu2020vehicular,xu2022wireless}.
        Distributed approaches eliminate intersection managers through direct CAV communication, including Virtual Traffic Light (VTL) systems with leader vehicles~\cite{ferreira2011impact,tonguz2013vehicle} and distributed consensus protocols using Perception-Initiative-Consensus-Action (PICA)~\cite{hildebrand2023comprehensive,regnath2021ciscav}.

        VTL elects a vehicle via V2V communication to temporarily control intersection traffic, demonstrating significant CO2 reduction and over 60\% traffic flow improvement without additional infrastructure costs~\cite{ferreira2011impact,ferreira2010self}. However, leader dependency creates vulnerabilities, including network failures, unexpected leader movement, and potential system-wide failure during malfunctions.

        Distributed consensus algorithms address node failure through Byzantine Fault Tolerance (BFT) for arbitrary malfunctions and Crash Fault Tolerance (CFT) for simple crashes~\cite{lamport2019part}. Current research explores blockchain-based approaches that involve PoW or PoS for block validation~\cite{king2012ppcoin,gervais2016security}. State machine replication techniques are also being studied, such as enhanced two-hop Raft frameworks~\cite{10000723} and PBFT protocols optimized for dynamic wireless environments~\cite{10041971}.
        
        \begin{table}[h!]
        \caption{\textbf{Comparison of Distributed Consensus Algorithms.} \\
        \cite{hildebrand2023comprehensive} has been reconstructed.}
        \label{tab:consensus_comparison}
        \centering
        \renewcommand{\arraystretch}{2}
        \resizebox{0.45\textwidth}{!}{
        \begin{tabular}{lcccc}
        \hline
        \textbf{Properties}              & \textbf{PoW}             & \textbf{PoS}          & \textbf{PBFT}         & \textbf{Raft}         \\ \hline
        Participation Cost               & Yes                      & Yes                   & No                    & No                    \\
        Trust Model                      & Untrusted                & Untrusted             & Semi-trusted          & Semi-trusted          \\
        Scalability                      & High                     & High                  & Low                   & Low                   \\ \hline
        Throughput                 & $<$10                    & $<$1,000              & $<$10,000             & $>$10,000             \\
        Byzantine Fault Tolerance        & 50\%                     & 50\%                  & 33\%                  & -                 \\
        Crash Fault Tolerance            & 50\%                     & 50\%                  & 33\%                  & 50\%                  \\
        Confirmation Time                & $>$100s                  & $<$100s               & $<$10s                & $<$10s                \\ \hline
        \end{tabular}
        }
        \end{table}
        
        However, PoW and PoS mechanisms face limitations in vehicular environments due to high computational complexity, low throughput, and excessive power consumption~\cite{10041971}. Additionally, PBFT and Raft have scalability constraints and require significant modifications for V2V wireless communication, having been designed for traditional wired networks. Therefore, new consensus mechanisms optimized for vehicular networks are needed, considering low latency, high throughput, and limited computing resources, along with lightweight algorithms ensuring reliability in unstable wireless environments.
        
        We aim to address the issue of determining the entry order in simultaneous entry or equal priority situations at intersections, and to achieve this, we propose a new approach based on the Raft algorithm, which follows a decentralized approach. Since Raft is originally optimized for wired network environments, only the core elements of the algorithm are selectively applied to make it suitable for dynamic wireless communication environments between vehicles. In particular, since the voting mechanism has been proven effective as a negotiation mechanism for autonomous vehicles in existing studies~\cite{boos2020networks, teixeira2018autonomous}, the voting mechanism and quorum concept of Raft are utilized as key elements.
        
        The proposed approach can resolve the shortcomings of the existing distributed consensus methods summarized in Table~\ref{tab:consensus_comparison}. We adapt Raft algorithm for high throughput and short confirmation time. We exclude high-uncertainty situations such as random selection that do not guarantee safety and liveness. Instead, we utilized a verified voting mechanism that ensures stable and reliable consensus.
\section{Consensus Algorithm Design}\label{consensus_algorithm_design}

In this section, we assume that CAVs are equipped with vision systems and present a solution that determines the passing order at unsignalized intersections during simultaneous entries using a quorum-driven, voting-based distributed consensus method.

In distributed consensus systems, failure types are generally classified as stopping faults where nodes completely stop or fail to respond and Byzantine faults where nodes behave maliciously or transmit incorrect information intentionally~\cite{lynch1996distributed}. 
In the intersection environment considered in this study, failures mainly arise from unresponsive scenarios and communication delays of CAVs and HVs, which correspond to stopping faults. To address these issues, we adopt the voting process of the Raft algorithm, enabling the system to maintain consensus and continue functioning even when some nodes become unresponsive~\cite{zhan2023improvement}.
Although BFT-type algorithms can handle Byzantine faults~\cite{castro2002practical}, they require at least $3f + 1$ processes where $f$ is the number of Byzantine processes. Furthermore, they also pose the risk of indefinite delays making them impractical for intersection scenarios. Therefore, this study focuses solely on stopping faults and does not consider Byzantine faults or malicious behavior.

This algorithm targets CAVs at Levels 4 and 5, based on the standards of the SAE. CAVs have the following capabilities.
\begin{enumerate}
\item {All vehicles are equipped with computer vision technologies, including vision sensors, radar, and LiDAR.}
\item{Vehicles are equipped with V2X communication technologies.}
\item {Vehicles can detect the presence and the order of surrounding vehicle arrivals at the intersection through their sensors.}
\item {In simultaneous entry scenarios, all vehicles perceive the situation consistently through their vision systems.}
\item {During pass-through of the unsignalized intersection, CAVs' sensory system fully functions to evade any collision with abruptly crossing HVs against the consensus on entry priority.}
\item{Each CAV exposes the intended movement direction to other CAVs at the intersection.}
\end{enumerate}

CAVs are sensitive to the risk of process failures within the network and the feasibility of reaching consensus. In case of a consensus failure due to excessive communication delays and losses, we assume CAVs to be capable of switching to their secondary artificial vision system for determining the entry priorities based on the lexicographical order of the license plate numbers. 

\subsection{Intersection Consensus Flow}
\begin{figure}[h]
    \centering
    \includegraphics[width=0.5\textwidth]{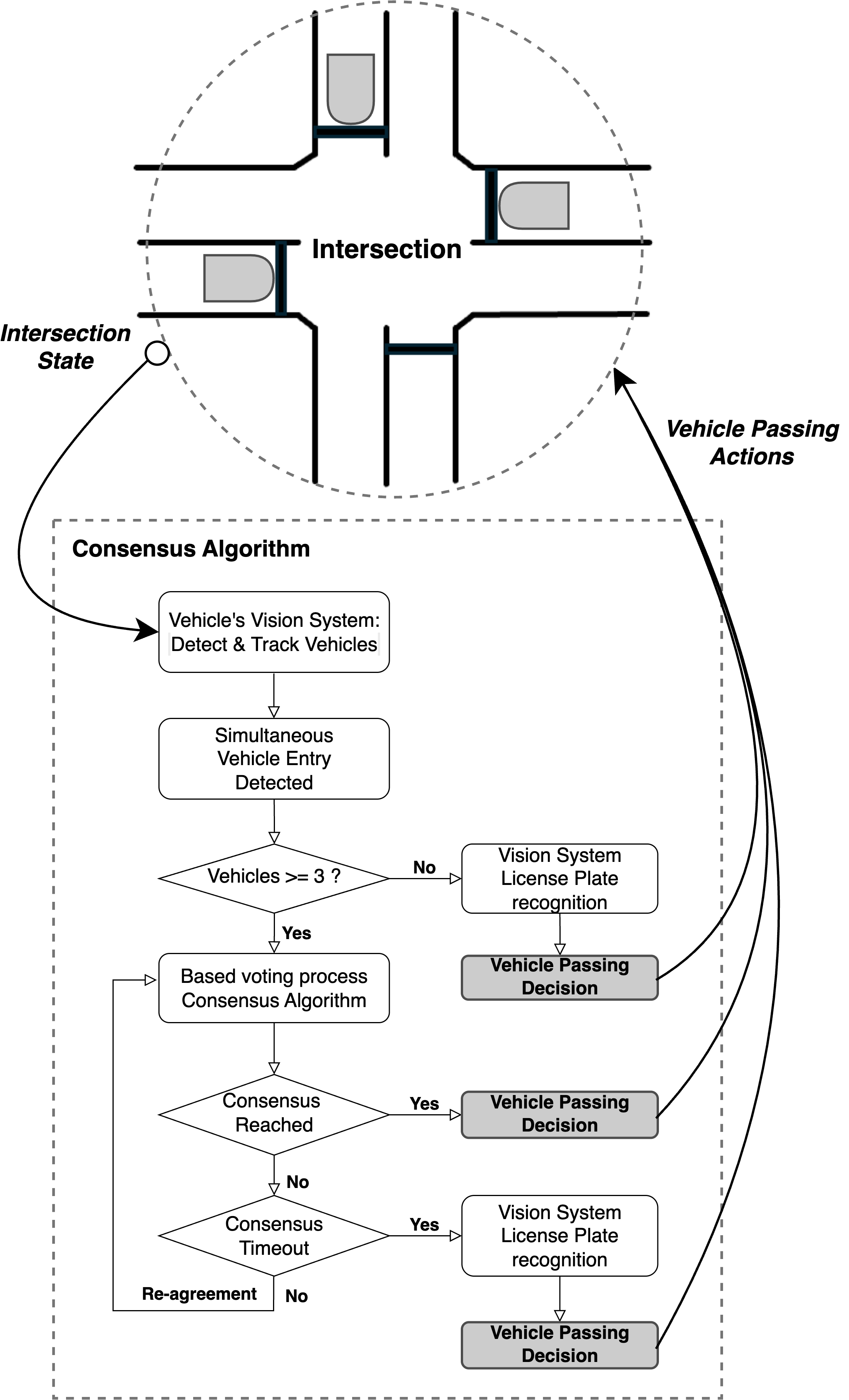}
    \caption{Intersection Consensus Flow Diagram : Vehicles entering simultaneously can detect how many other vehicles are waiting to enter the intersection upon their arrival. If the number of waiting vehicles is three or more, a voting-based consensus algorithm is triggered.}
    \label{fig:consensus_flow}
\end{figure}

The overall flow of the algorithm at the intersection is shown in Fig.\ref{fig:consensus_flow}.
This study proposes a novel system for achieving intersection crossing consensus among CAVs. The proposed system utilizes the vehicles’ vision systems to detect the presence of simultaneously entering vehicles and determines an appropriate consensus method based on the number of vehicles involved. 

\subsubsection{{Vehicle Count-Based Consensus}}
When there are three or more vehicles, a voting-based consensus algorithm is used. Consensus is achieved only when a majority of votes is obtained. When there are two or fewer vehicles, consensus through voting is not possible, which is similar to the behavior of Raft under quorum requirements. In particular, two vehicles cannot reach consensus through asynchronous communication with the arrival of messages being non-deterministic. This problem of never reaching consensus in an unreliable network channel is known as the Two Generals' Paradox~\cite{gray1978notes}. In such cases, a vision system that recognizes license plates comes into play to determine the crossing order to ensure the liveness of our consensus system. 

\subsubsection{{Consensus Cycle and Re-consensus}}
The consensus cycle is considered complete when at least one vehicle has passed through the intersection, which can be verified via the vision system. If no vehicle passes due to a failure to reach consensus, a re-consensus process is initiated. During this process, if the time taken for voting-based consensus exceeds a threshold due to an unsatisfied quorum requirement, the system automatically switches to a license plate recognition-based method. The timeout is defined as the maximum duration required by the vision system to determine the passing order. In the vision-based decision mode, crossing priority is set according to the lexicographical order.

\subsection{Voting-based Consensus Model}
\label{sec:voting_based_consensus_model}
This study proposes a voting-based distributed consensus algorithm that partially incorporates the principles of the Raft algorithm to determine the passing priority among CAVs at unsignalized intersections. Each vehicle starts in the Candidate state. Upon receiving votes exceeding the quorum threshold, one vehicle is ultimately elected as the Leader. The proposed approach facilitates a more efficient and fair decision-making process for intersection crossing.

\renewcommand{\arraystretch}{2}
\begin{table}[H]
\caption{Algorithm Variables maintained at Each Participating Vehicle}\label{tab:AlgorithmVariables}
    \centering
    \fontsize{15}{13}\selectfont
    \resizebox{0.45\textwidth}{!}{
    \begin{tabular}{|c|p{10cm}|}
        \hline
        \textbf{Variable} & \textbf{Description} \\ \hline
        $N_n$ & Network address of vehicle $N$ \\ \hline
        $D_n$ & Path direction of vehicle $N$ \\ \hline
        sent-votes & Number of votes sent by the vehicle \\ \hline
        received-votes & Number of votes received by the vehicle \\ \hline
        no-collision-list & List of vehicles that are not on a collision path \\ \hline
        election-status & The state during the leader election voting process. It indicates the current role or status of the vehicle in the election, such as whether it is a "Candidate," "Follower," or "Leader" \\ \hline
        election-time & Time when the leader decision was made (last response time recorded) \\ \hline
        election-received-votes & Number of votes received during the election process \\ \hline
        $T_{m \to n}$ & Time when vehicle $N$ received a message from vehicle $M$ (recorded in election-time) \\ \hline
        Quorum & The minimum number of votes required to achieve consensus, typically a majority (more than half) of the total number of vehicles \\ \hline
    \end{tabular}
    }
\end{table}
In this section, we describe the algorithm variables and operational principles of the proposed voting-based algorithm.
This algorithm requires the variables listed in Table \ref{tab:AlgorithmVariables}. The proposed voting-based consensus model operates through two voting phases. During each phase, vehicles interact by exchanging messages, as detailed in Algorithm \ref{tab:Response for Candidate Vote Request} and Algorithm \ref{tab:Response for Leader Election Vote Request}. Since the Candidate status appears in both rounds of the voting process, we distinguish them as $Init\text{-}Candidate$ and $Fin\text{-}Candidate$ in the election-status.

\subsubsection{{Candidate Voting Process}}
Vehicles that arrive at the intersection simultaneously are considered to have entered within the same time window. Since all vehicles initiate the process together, they start as Candidate—without a preceding Follower state, unlike the standard Raft protocol. We refer  to this initial state as $Init\text{-}Candidate$. Each vehicle sends a vote request to all other vehicles except itself in an attempt to become the $Leader$. At the same time, each vehicle responds to incoming vote requests from others. Only the request received first can be acknowledged. Each vehicle is allowed to vote only once. If a vehicle receives an "acknowledged" response, it compares the direction of the sender, the acknowledging vehicle, with its own to assess the possibility of the sender vehicle being on the collision path. The vehicle adds the sender’s address to its no-collision-list if it is not on the collision path and updates its election-time to the time the acknowledgment was received.

\begin{algorithm}[h!]
\caption{Response for Candidate Vote Request}
\label{tab:Response for Candidate Vote Request}
\begin{algorithmic}[1]
\State \textbf{lock} access to shared resource (using mutex)
\If{request is null}
    \State \textbf{unlock} access to shared resource
    \State \textbf{return} error: "received nil request"
\EndIf
\If{current vehicle has not sent any votes yet}
    \State mark the current vehicle as having voted (set sent-votes to 1)
    \If{direction in valid\_directions}
        \State directionStatus $\gets$ $True$
    \Else
        \State directionStatus $\gets$ $False$
    \EndIf
    \State response $\gets$ prepareResponse("Acknowledged", directionStatus, Vehicle)
\Else
    \State response $\gets$ prepareResponse("Ignored", Vehicle)
\EndIf
\State \textbf{unlock} access to shared resource
\State \textbf{return} response
\end{algorithmic}
\end{algorithm}

\subsubsection{{Leader Election Voting Process}}
If an $Init\text{-}Candidate$ collects votes exceeding the quorum, it proceeds to initiate a Leader Election by sending a Leader Election Request message to all other vehicles. At this point, the Candidate is referred to as a $Fin\text{-}Candidate$. Upon receiving this Leader Election Request, each recipient vehicle compares its own vote count with that of the $Fin\text{-}Candidate$ (the sender of the request). If the $Fin\text{-}Candidate$ has more votes, the recipient sends an "acknowledged" response; otherwise, it replies with "ignored." If both vote counts are equal, the recipient compares election-time and acknowledges the vehicle that reached quorum earlier.

When a vehicle sends an “acknowledged” response to the $Fin\text{-}Candidate$, it transitions from $Init\text{-}Candidate$ to $Follower$ or $Fin\text{-}Candidate$ to $Follower$. It also updates its election-time and received-votes to match those of the $Fin\text{-}Candidate$. This mechanism ensures that any future requests from a vehicle with fewer votes or a later election-time will not override the current $Leader$. Conversely, if the recipient vehicle sends an “ignored” response, the $Fin\text{-}Candidate$ itself transitions to $Follower$, updating its own election-time and received-votes to match those of the sender (the vehicle that sent the “ignored” response). Once a vehicle transitions to $Follower$, it becomes ineligible to participate in the ongoing Leader Election Voting Process.

\begin{algorithm}[h!]
\caption{Response for Leader Election Vote Request}
\label{tab:Response for Leader Election Vote Request}
\begin{algorithmic}[1]
    \If {req.Vehicle.election-status is $Follower$}
        \State \Return ignored
    \EndIf
    \If {s.Vehicle.received-votes \textless req.Vehicle.received-votes \textbf{or} 
         (s.Vehicle.received-votes = req.Vehicle.received-votes \textbf{and} req.Vehicle.election-time \textgreater s.Vehicle.election-time)}
        \State s.Vehicle.election-status $\gets$ $Follower$
        \State s.Vehicle $\gets$ req.Vehicle
        \State \Return acknowledged
    \Else
        \State req.Vehicle.election-status $\gets$ $Follower$
        \State req.Vehicle $\gets$ s.Vehicle
        \State \Return ignored
    \EndIf
\end{algorithmic}
\end{algorithm}

A $Fin\text{-}Candidate$ that receives “acknowledged” responses from a majority of vehicles is confirmed as the $Leader$ for the current consensus term. The elected “Leader” is granted the right to pass through the intersection, and all participating vehicles confirm this event via their vision system. This ensures both a clear completion point for the consensus cycle and verifiable passage of the elected “Leader”.

\subsection{Distributed Consensus Scenario}
\begin{figure}[h!]
\centering
\includegraphics[width=0.39\textwidth]{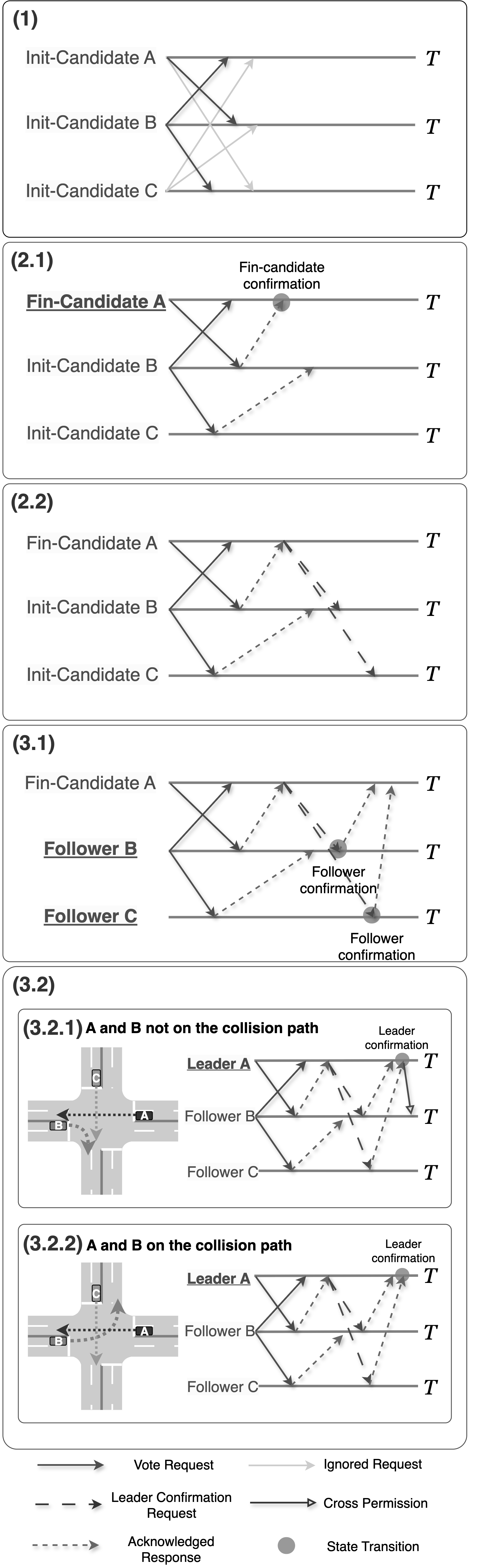}
\caption{The consensus process among CAVs simultaneously entering an unsignalized intersection; $T$ is the consensus timeout.}
\label{fig:ConsensusProcedure}
\end{figure}

Based on the proposed voting algorithm, we explain step-by-step message exchange procedure in a scenario where three CAVs simultaneously arrive at the intersection as illustrated in Fig.~\ref{fig:ConsensusProcedure}.

\subsubsection{Initial State at Unsignalized Intersections}
All vehicles entering an unsignalized intersection simultaneously start in the $Init\text{-}Candidate$ state unlike the original Raft algorithm where all processes start in the $Follower$ state. Each vehicle initializes its address and direction and first casts a vote for itself. Subsequently, it simultaneously sends vote requests to all other vehicles and performs the Candidate Voting Process as described in Section~\ref{sec:voting_based_consensus_model}.

Fig.~\ref{fig:ConsensusProcedure} (1) illustrates the process in which vehicles $A$, $B$, and $C$, in the $Init\text{-}Candidate$ state, simultaneously send vote requests to one another. Votes are cast only for the first received request, with all subsequent requests being rejected.

\subsubsection{Candidate Voting Response and State Transition}
Vote responses are classified into two types, which are "acknowledged" and "ignored". Each response includes information about whether a collision occurs and the vehicle's direction. In Fig.\ref{fig:ConsensusProcedure} (2.1), vehicle $A$ receives an "acknowledged" response from vehicle $B$. If $B$'s direction does not collide with that of $A$, $B$'s address is added to $A$'s no-collision-list.
Notably, $A$ secures a majority of votes, including its own, and transitions to the $Fin\text{-}Candidate~A$ state. Subsequently, it proceeds to the Leader Election Voting Process and sends leader election vote requests to all other vehicles (as shown in Fig.\ref{fig:ConsensusProcedure} (2.2)).

\subsubsection{Leader Election Voting Response and State Transition}
Vehicles receiving requests from the $Fin\text{-}Candidate$ evaluate whether to agree to or refuse the leader election based on their own received-votes and the Fin-Candidate election-time of the last vote received (as shown in line 4 of Algorithm 2). In Fig.\ref{fig:ConsensusProcedure} (3.1), vehicle $A$ has a higher vote count and an earlier election-time than vehicles $B$ and $C$. Therefore, $A$ receives "acknowledged" responses from both $B$ and $C$. During this process, vehicles $B$ and $C$ transition to the $Follower$ state, while $A$, having achieved quorum, is ultimately elected as the $Leader$ and is granted priority to proceed through the intersection. The times when a first vote request is issued by each vehicle to the neighboring vehicles are randomly varied by the Raft algorithm. This simple variation makes it highly unlikely to have more than one CAV in $Fin\text{-}Candidate$ state to get the same vote count at the same time, ultimately producing only one leader vehicle.

If there is no directional collision between $A$ and $B$, and $B$'s address has been added to $A$'s no-collision-list (as shown in Fig.\ref{fig:ConsensusProcedure} (3.2.1)), then $A$ sends a message that permits $B$ to pass through the intersection. Conversely, if $B$ does not exist in the no-collision-list (Fig.\ref{fig:ConsensusProcedure} (3.2.2)), $A$ does not transmit such a message.

\subsubsection{Termination of the Consensus Cycle and Re-consensus}


The consensus cycle is considered complete once the elected Leader vehicle successfully passes through the intersection. If not all vehicles that entered the intersection simultaneously have passed, the remaining vehicles restart the consensus procedure from the beginning.

However, if the Leader election times out, the participating vehicles automatically switch to a vision-based priority decision method to avoid further delays. The condition for this transition is defined as follows:
\begin{equation}
    \text{If } T_{\text{consensus}} \geq T_{\text{vision}} \\
    \Rightarrow~\text{Use Vision-Based Decision}
\end{equation}
Here, \( T_{\text{consensus}} \) represents the time taken by the voting-based consensus process, and \( T_{\text{vision}} \) denotes the threshold time required for priority determination using the vision system.

\subsection{Analysis of Safety and Liveness}
The proposed voting-based consensus algorithm and experiments are conducted in an asynchronous environment where message delays are non-deterministic. The FLP impossibility proof demonstrates that in an asynchronous system, no deterministic protocol can guarantee consensus among reliable processes without the possibility of nontermination, even if there is only a single crash-prone process~\cite{fischer1985impossibility}. However, by setting the quorum threshold to a strict majority, the proposed algorithm practically derives an optimal quorum condition that enables consensus among correctly responding CAVs while satisfying both properties.

\subsubsection{Proof of Safety}
First, let the number of vehicles (either CAVs or HVs) be denoted as $V$, and assume that the quorum is set to a value less than the majority. This assumption is used to construct a proof by contradiction, showing that setting the quorum to less than a majority cannot guarantee safety.

Let the quorum threshold be $k$, where $(k < \frac{V}{2})$, and define the set of all vehicles as $S$, with $\vert S \vert = V$. Now suppose that two disjoint subsets of vehicles, $A$, $B$, each satisfy the quorum condition, i.e.,
\begin{equation}
    \vert A \vert \geq k \quad \text{and} \quad \vert B \vert \geq k
\end{equation}
If subsets $A$ and $B$ are disjoint, then:
\begin{equation}
    \vert A \vert + \vert B \vert \leq V
\end{equation}
This implies that both sets can satisfy the quorum condition simultaneously. If set $A$ reaches consensus on value $v_1$ and set $B$ on value $v_2$, where $v_1 \neq v_2$, it is possible for two different values to be agreed upon at the same time. This leads to a violation of the safety property. The situation can be formally described as follows:
\begin{equation}
    \vert A \vert \geq k, \quad \vert B \vert \geq k, \quad \vert A \vert + \vert B \vert \leq V, \quad v_1 \neq v_2
\end{equation}
Therefore, if the quorum is set to less than a majority, it is possible for multiple disjoint subsets to reach consensus on different values simultaneously, thereby violating the safety property. This contradiction proves that the quorum must be greater than or equal to the majority to ensure the safety property. Furthermore, under this majority condition, it is guaranteed that only one Fin-Candidate can ultimately be elected as the Leader. 

\subsubsection{Guarantee of Liveness}
The proposed algorithm guarantees liveness under the condition that up to $f$ HVs or unresponsive autonomous vehicles are present. As long as $f+1$ responses are received, a majority can be formed, enabling consensus to be reached. This guarantee aligns with the quorum requirements established in Raft~\cite{ongaro2014search}. 
For example, when the total number of vehicles is $V=2f+1$, the majority threshold is $f+1$. Thus, even if up to $f$ messages are not received, liveness can still be preserved. This implies that consensus can be achieved solely among the responsive vehicles, even when some CAVs fail to respond due to system faults, network delays, or message losses, or when HVs are inherently unresponsive. Although unlikely, if all vehicles simultaneously become Fin-Candidates and the votes are split, no candidate would receive a majority, triggering a re-consensus. This process repeats until a Leader emerges. To avoid an infinite re-consensus cycle, the algorithm incorporates a timeout mechanism.

\subsubsection{Optimal Quorum for Safety and Liveness Guarantee}
In our study, a crashed process refers to either an HV or an unresponsive CAV. In other words, if HVs or unresponsive CAVs mixed in the intersection are not excluded from the consensus process, agreement may never be reached.

Therefore, a practical solution is to ensure that consensus can be achieved at least among the responsive CAVs within the quorum. In this study, we set the minimum quorum as $k > \left\lceil \frac{N}{2} \right\rceil$, allowing safe agreement and liveness among the normally responding vehicles. Setting this minimum quorum is essential to enable rapid consensus among responsive CAVs and prevent congestion at intersections. The effectiveness of this quorum threshold in intersections where HVs and unresponsive autonomous vehicles coexist is evaluated through simulation experiments in the following Section \ref{implementation_evaluation}.
\section{implementation and Evaluation}\label{implementation_evaluation}


This section presents the implementation and evaluation of the proposed voting algorithm using gRPC over an Ethernet emulating lossy VANET. The algorithm’s efficiency and responsiveness are assessed using performance indicators-throughput and consensus time. Ultimately, we demonstrate that the algorithm satisfies both liveness and safety properties while also identifying the most optimized intersection scenario.

gRPC~\cite{indrasiri2020grpc}, an open-source RPC framework released by Google in 2015, is optimized for rapid consensus and real-time processing. It achieves this by combining high-speed data transmission over HTTP/2 with lightweight serialization through Protocol Buffers. Its support for multiplexing and streaming minimizes latency while ensuring cross-language compatibility, making it well-suited for distributed systems. Leveraging these advantages, we designed a communication architecture tailored to intersection environments that require rapid consensus. 

Given that concurrency and asynchrony are central to the proposed algorithm, the implementation was carried out in Go, which supports lightweight thread models (goroutines) and efficient asynchronous processing. These features make it particularly suitable for achieving fast consensus in intersection scenarios. The Go implementation can later be converted to an embedded C/C++ firmware for a deployment to the fleet of actual CAVs.  

\subsection{Experimental Design}



The experiments were conducted with two main variables—the number of vehicles participating in the consensus process and the HV-CAV ratio. The time indicators used in the experiments were designed with reference to related literature, as some metrics were difficult to calculate directly or accurately reflect in practice. For autonomous vehicles, it is challenging to precisely measure the time required to recognize and compare license plates. However, based on results combining Optical Character Recognition (OCR) with YOLOv5, the average processing time per image on a standard desktop computer was approximately 0.5 seconds~\cite{Kala2024AutomaticNP}. Accordingly, this study sets the processing time per vehicle for the vision system as 0.5 seconds, defined as $T_{vision}$, and assumes that vehicles can recognize multiple license plates simultaneously in parallel.


In contrast to autonomous vehicles, HVs need additional time to perceive the situation and adjust their speed. Empirical studies report reaction times of about 1.2-1.5 seconds for younger adults and up to 2.5 seconds for older drivers in complex situations such as uncontrolled intersections or yielding~\cite{FHWA_RD_97_135}. To model a best-case, responsive human driver and to avoid overestimating the benefit of CAVs, we set the decision-making time of HVs to the lower bound of 1.2 seconds in our experiments.

In this experiment, a consensus timeout refers to a situation where the voting-based consensus time ($T_{\text{consensus}}$) exceeds the vision-based threshold ($T_{\text{vision}}$), requiring the vision system to take over complete the consensus process. Experimental results
show that the proposed algorithm reached consensus at a typical unsignalized four-way, two-lane intersection in approximately 30-40 ms on average. 

\subsection{Baseline Comparison}
The proposed model was evaluated against conventional traffic signal control and centralized AIM in terms of decision latency and deployment cost. The experiment assumes an unsignalized four-way, two-lane intersection in which four vehicles enter simultaneously (Table.~\ref{tab:decision_time}). 

Conventional signal control inherently produces fixed cycle delays regardless of the CAV penetration rate. Although typical cycle lengths range from 45 to 120 seconds~\cite{PennDOT_2021}, we adopted 45 seconds as a lower-bound baseline to represent the fastest possible response scenario. Centralized AIM achieved a decision latency of approximately 4.7 milliseconds under a 100\% CAV condition, while the proposed model recorded 30 milliseconds under the same condition. Although centralized AIM remains faster in the ideal setting, the proposed model still demonstrates a substantial latency reduction compared to conventional traffic signal control. In the 0\% CAV scenario, V2X negotiation does not occur, and therefore both centralized AIM and the proposed model become inoperative and equivalent to a conventional unsignalized intersection.
\begin{table}[H]
\centering
\caption{Decision latency comparison}
\label{tab:decision_time}
\fontsize{15}{13}\selectfont
\resizebox{0.48\textwidth}{!}{%
    \begin{tabular}{lcccc}
        \hline
        \textbf{Model} & \textbf{100\% CAV} & \textbf{75\% CAV} & \textbf{50\% CAV} & \textbf{0\% CAV} \\
        \hline
        Conventional Traffic Signal & 45 s & 45 s & 45 s & 45 s \\
        Centralized AIM             & 4.7 ms & 1.20 s & 1.204 s & N/A\textsuperscript{*} \\
        \rowcolor{gray!10} Proposed Model
        & 30 ms & 1.72 s & 2.9 s & N/A\textsuperscript{*} \\
        \hline
    \end{tabular}%
}

\vspace{2pt}
\footnotesize\textsuperscript{*}\,Not applicable because no CAV communication is available.
\end{table}

In terms of deployment cost (Table.~\ref{tab:cost_comparison}), installing a conventional fixed-signal system requires approximately 200{,}000 USD per intersection~\cite{clarksville_ss4a_2024}. Centralized AIM requires roadside units (RSUs) and V2I communication equipment, as well as C-V2X on-board units (OBUs) that add roughly 160–170 USD per vehicle~\cite{itsamerica_v2x_2023}. In contrast, the proposed model does not rely on RSUs or other fixed roadside infrastructure and operates solely through OBU-based distributed negotiation, thereby achieving substantially lower deployment cost than centralized AIM while retaining the latency advantages over conventional signal control.
\begin{table}[H]
\centering
\caption{Deployment Cost comparison}
\label{tab:cost_comparison}
\fontsize{15}{13}\selectfont
\resizebox{0.48\textwidth}{!}{%
    \begin{tabular}{lcc}
        \hline
        \textbf{Model} & \textbf{Infrastructure cost} & \textbf{Vehicle OBU cost} \\
        \hline
        Conventional Traffic Signal & 200,000 USD & 0 USD \\
        Centralized AIM             & 50,000 USD & 160--170 USD \\
        \rowcolor{gray!10} Proposed Model & 0 USD & 160--170 USD \\
        \hline
    \end{tabular}%
}
\end{table}

\subsection{Distributed AIM Comparison}


We directly compare the consensus latency characteristics of PBFT—one of the most widely used distributed consensus algorithms—and the proposed model. The experiment assumes an unsignalized four-way intersection where four vehicles enter simultaneously, and evaluates each algorithm in terms of consensus time and timeout rate. PBFT guarantees Byzantine fault tolerance under the $3f+1$ process configuration. However, when this scheme is applied to a small intersection group involving only three to four vehicles, its quorum requirement becomes structurally restrictive. In particular, when only three vehicles participate, the $3f+1$ condition cannot be satisfied, making consensus theoretically unattainable. Considering that non-responsive vehicles are more likely under low CAV penetration, timeout termination becomes even more frequent.

In contrast, the proposed model adopts a $2f+1$ majority requirement that is better suited for small consensus groups at unsignalized intersections, providing structural advantages in scenarios with limited participants. Furthermore, even under ideal conditions where all vehicles respond and both models can theoretically satisfy quorum requirements, the proposed method reaches consensus more quickly than the PBFT-based baseline. As shown in Table.~\ref{tab:consensus_comparison}, the proposed model completes consensus faster in the 100\% CAV scenario, and maintains stable consensus without timeout under 75\% and 50\% CAV penetration. Meanwhile, the PBFT-based model exhibits increased delay and significantly higher timeout rates under the same conditions.

\begin{table}[H]
\centering
\caption{Consensus performance comparison}
\label{tab:consensus_comparison}
\fontsize{15}{13}\selectfont
\resizebox{0.48\textwidth}{!}{%
    \begin{tabular}{lcccc}
        \hline
        \textbf{Model}
        & \textbf{Metric} & \textbf{100\% CAV} & \textbf{75\% CAV} & \textbf{50\% CAV} \\
        \hline
        \multirow{2}{*}{PBFT-based Model (3$f$ + 1)}
            & Consensus time    & 69.7 ms & 4.71 s & Fails to terminate \\
            & Consensus timeout & 0\%     & 34\%     & 100\%\textsuperscript{*} \\
        \hline
        \multirow{2}{*}{Proposed Model (2$f$ + 1)}
            & Consensus time    & 30 ms   & 1.72 s  & 2.9 s \\
            & Consensus timeout & 0\%     & 0\%     & 0\% \\
        \hline
    \end{tabular}%
}

\vspace{2pt}
\footnotesize\textsuperscript{*}\,Timeout occurs in all trials because the $3f+1$ quorum requirement cannot be satisfied.
\end{table}

\subsection{Consensus Quorum}
\begin{figure}[h!]
    \centering \includegraphics[width=0.48\textwidth]{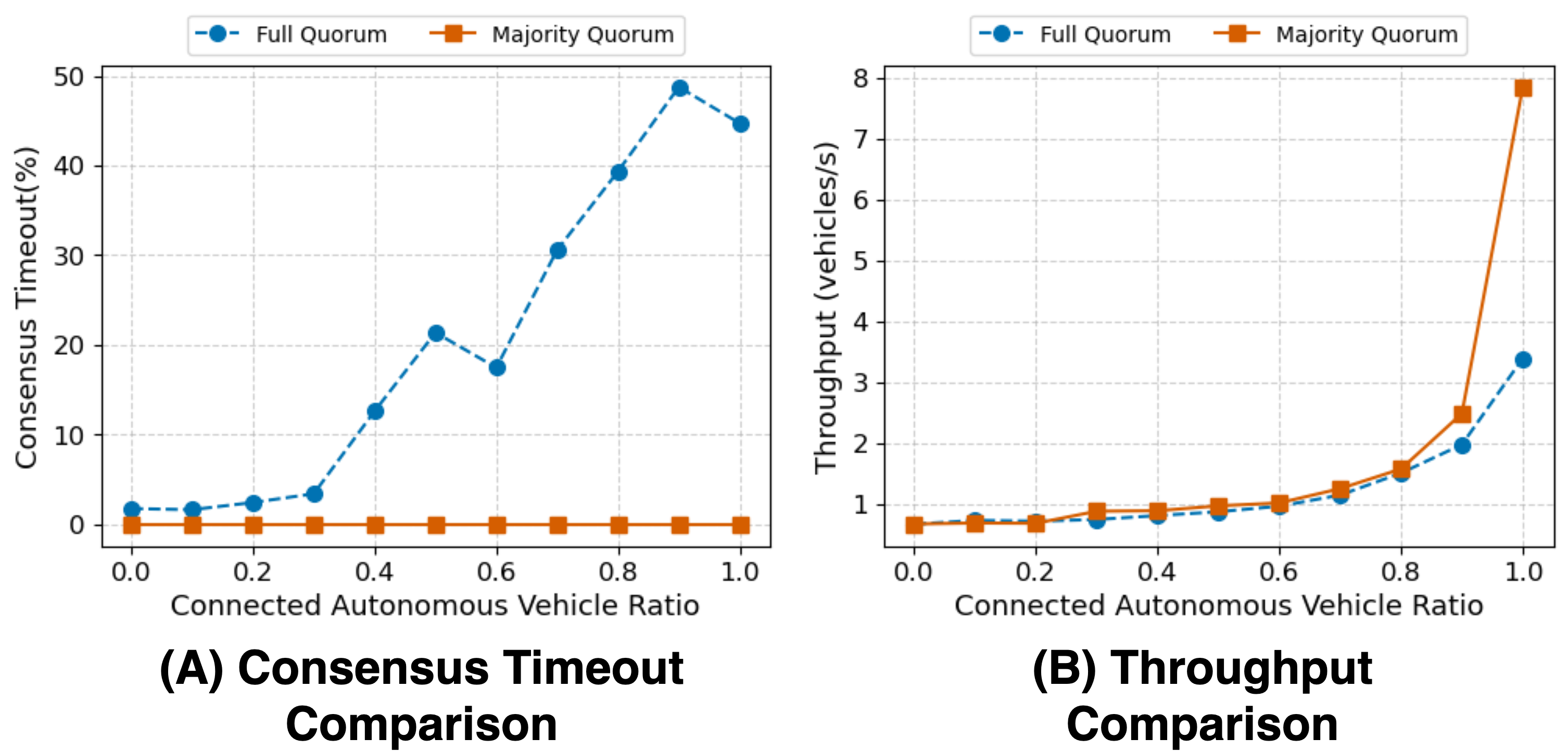
    }
    \caption{Comparison of consensus timeout rates and throughput under Full Quorum and Majority Quorum conditions according to varying CAV ratios.}
    \label{fig:quorum_comparison}
\end{figure}

To demonstrate the efficiency and stability of consensus when the quorum is set to a majority, experiments were conducted by comparing scenarios where the quorum was set to a majority quorum and a full quorum (Fig.~\ref{fig:quorum_comparison}). The experiments were performed at a four-way intersection with two lanes, involving 300 vehicles under identical conditions. The results showed that setting the quorum to a majority quorum provided clear advantages over the full quorum during the consensus process. With the majority quorum, there were no cases where the consensus time exceeded $T_{vision}$, even in the presence of unresponsive vehicles or HVs. In contrast, with the full quorum, the timeout rate approached 50\% as the proportion of CAVs increased, resulting in greater reliance on vision-based technology. In terms of throughput, the majority quorum also demonstrated significantly better performance than the full quorum, with the difference becoming particularly pronounced when the CAV ratio exceeded 80\%. These experimental results show that setting the quorum to a majority can reduce reliance on the vision system while improving system throughput. This logically supports the conclusion that a majority quorum represents the optimal quorum condition in practice.

\subsection{Scenarios}

The performance of the proposed algorithm can vary depending on different intersection environments and operating conditions. Therefore, we systematically evaluated its efficiency by adjusting key variables such as intersection size, $T_{vision}$, and the proportion of CAV.

\begin{figure}[h!]
    \centering \includegraphics[width=0.48\textwidth]{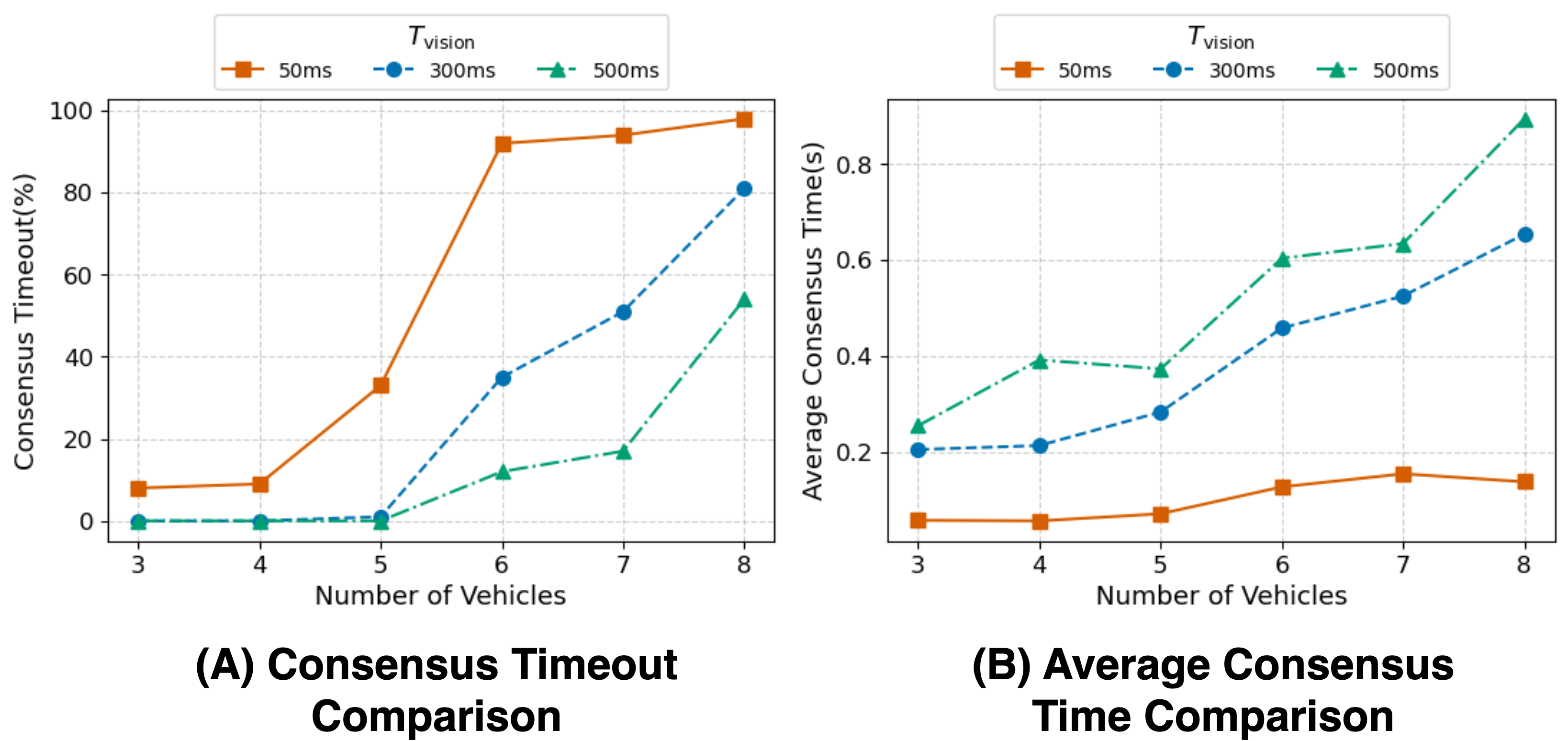
    }
    \caption{Comparison of consensus timeout rates and average consensus times under different $T_{\text{vision}}$ values as the number of vehicles increases.}
    \label{fig:Timeout Comparison}
\end{figure}

In the first experiment, we analyzed the algorithm’s performance with respect to variations in $T_{vision}$ (Fig.~\ref{fig:Timeout Comparison}). While the baseline experiments fixed $T_{vision}$ at 500 ms, additional tests were conducted with $T_{vision}$ set to 50 ms and 300 ms to account for potential reductions in license plate recognition and perception time resulting from future technological advancements. The results showed that shorter $T_{vision}$ values generally led to reduced average consensus times. However, as the number of vehicles increased, it became increasingly difficult to achieve consensus within $T_{vision}$, leading to a higher proportion of cases exceeding this threshold. This trend was particularly pronounced when five or more vehicles entered the intersection simultaneously.

\begin{figure}[h!]
    \centering \includegraphics[width=0.48\textwidth]{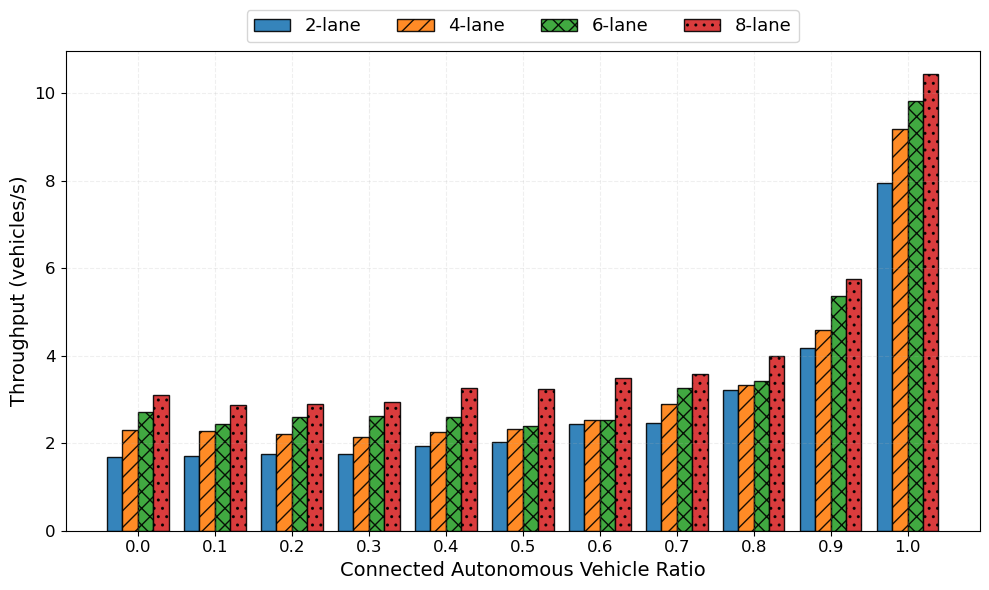}
    \caption{Comparison of throughput across different lane configurations as the CAV ratio increases.}
    \label{fig:Lane Comparison}
\end{figure}

In the second experiment, we conducted a comprehensive analysis of the algorithm’s performance under varying intersection capacity and CAV ratios (Fig.~\ref{fig:Lane Comparison}). To precisely analyze scenarios in unsignalized intersections with mixed CAV and HV traffic, we conducted experiments with 300 consecutively passing vehicles, systematically varying the CAV ratio from 0\% to 100\%. For experimental accuracy, variables such as vehicle speed and inter-vehicle distance were controlled, and only the decision-making time during simultaneous entries was measured. The experiments were performed under four intersection configurations to reflect different intersection sizes: two-lane, four-lane, six-lane, and eight-lane four-way intersections.

The experimental results showed that when the CAV ratio exceeded 60\%, throughput consistently increased across all lane configurations, with the rate of increase becoming more pronounced as the number of lanes grew.

\section{Discussion and Conclusion}\label{conclusion}  


We proposed a distributed consensus algorithm that applies voting-based principles of Raft to achieve fast and reliable agreement among vehicles even during simultaneous entries at unsignalized intersections.
The baseline comparison showed that the proposed model reduces decision latency without requiring additional infrastructure. In small intersection groups, the distributed AIM comparison further confirmed that it reaches consensus faster than the PBFT-based model and incurs no timeouts under the same conditions.
Through a majority quorum design, the algorithm demonstrated superior performance in terms of throughput and timeout ratio compared to a full quorum. It also successfully achieved stable consensus within $T_{\text{vision}}$ even in environments where faulty CAVs and unresponsive HVs coexist, confirming its practical applicability. With a lower $T_{\text{vision}}$, we observed reduced overall average consensus despite a higher rate of consensus timeouts since the secondary vision system augmented the decision process based on the lexicographical order of license plate numbers.   
The proposed distributed algorithm showed the potential for deployment to unsignalized intersections by achieving high throughput and rapid consensus without a costly centralized management infrastructure. 

However,  several limitations remain. The study did not consider dynamic driving characteristics such as vehicle speed, acceleration, or braking distance, nor did it address complex traffic or environmental variables such as adverse weather conditions, nighttime scenarios, or pedestrian presence. 
Future work will aim to enhance the practical applicability of the algorithm by developing a more realistic simulation environment, for example by integrating microscopic traffic simulators such as SUMO, and by incorporating dynamic variables and real-world factors (e.g., pedestrians, bicycles, road obstacles, and adverse weather) into advanced experiments under diverse urban intersection types and environmental conditions.
Furthermore, to improve the adaptability and practical robustness of the algorithm in complex real-world scenarios, future research will focus on extending the current protocol framework to address Byzantine faults caused by compromised or malicious vehicles.


%



\section*{Acknowledgment}
This work was partly supported by the Institute of Information \& Communications Technology Planning \& Evaluation(IITP)-ITRC(Information Technology Research Center) grant funded by the Korea government(MSIT) (IITP-2026-RS-2023-00259099) and by 2026 Hongik University Research Fund.

\ifCLASSOPTIONcaptionsoff
  \newpage
\fi




\bibliographystyle{IEEEtran}

\bibliography{reference}

\begin{IEEEbiography}
[{\includegraphics[width=1in,height=1.25in,clip,keepaspectratio]{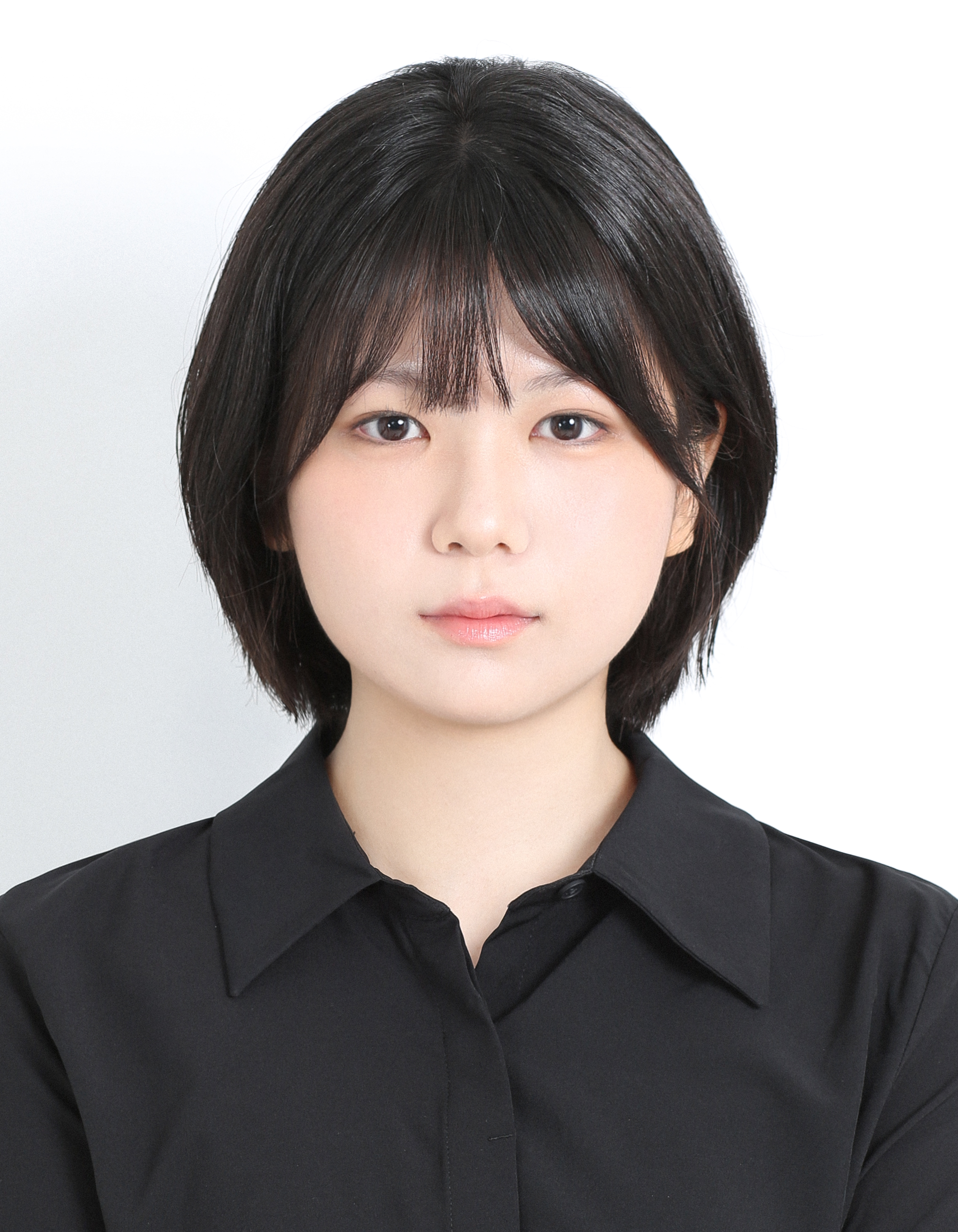}}]{Younjeong Lee}
is a research staff at Distributed Intelligence and Autonomy Lab in Hongik University with double majors in Urban Design \& Planning and Computer Engineering. Lee's research interest is in AI-based traffic management AI and large-scale distributed computing technologies. 
\end{IEEEbiography}
\begin{IEEEbiography}[{\includegraphics[width=1in,height=1.25in,clip,keepaspectratio]{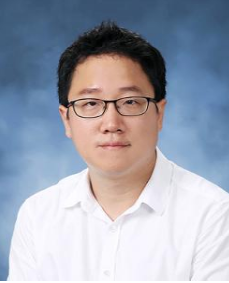}}]{Young Yoon} is an associate professor in computer engineering at Hongik University and a CTO for Neouly Incorporated. His research interest is in distributed intelligence, middleware, and cyber security. He recently published research articles on traffic analysis, public transportation management and autonomous vehicle accident studies using deep learning technologies. Yoon earned a BA and MS in computer sciences at the University of Texas at Austin in 2003 and 2006, respectively. He earned his Ph.D. degree in computer engineering at University of Toronto in 2013.
\end{IEEEbiography}





\end{document}